\documentclass[twocolumn,showpacs,preprintnumbers,amsmath,amssymb]{revtex4}

\usepackage{graphicx}
\usepackage{dcolumn}
\usepackage{bm}
\usepackage[latin2]{inputenc}

\begin{document}


\title{Decay dynamics of neutral and charged excitonic complexes in single InAs/GaAs quantum dots}

\author{M. Feucker}
\author{R. Seguin}
\author{S. Rodt}
 \email{srodt@sol.physik.tu-berlin.de}
\author{A. Hoffmann}
\author{D. Bimberg}
\affiliation{Institut für Festkörperphysik, Technische Universität Berlin, D-10623 Berlin, Germany}


\date{\today}

\begin{abstract}
Systematic time-resolved measurements on neutral and charged excitonic complexes (X, XX, X+, and XX+) of 26 different single InAs/GaAs quantum dots are reported. The ratios of the decay times are discussed in terms of the number of transition channels determined by the excitonic fine structure and a specific transition time for each channel. The measured ratio for the neutral complexes is 1.7 deviating from the theoretically predicted value of 2. A ratio of 1.5 for the positively charged exciton and biexciton decay time is predicted and exactly matched by the measured ratio indicating identical specific transition times for the transition channels involved.
 \end{abstract}

\pacs{Valid PACS appear here}
\maketitle

Single quantum dots (QDs) provide the key for quantum computing \cite{chen02, bianucci04} and quantum cryptography \cite{benson00}. For real devices, the dynamics of the recombination is of utmost importance, because it limits the maximum modulation frequency of the device and the coherence time of the emitted light. The electron and hole wave functions and, therefore, the oscillator strength varies from dot to dot because of structural variations \cite{schliwa07}. Therefore, a spread of decay times is expected for different QDs. Additionally, different excitonic complexes in the same QD show different decay times for two main reasons: First, the number of possible recombination channels varies due to differences of the electronic fine structure of the various complexes. Second, the oscillator strength of a specific transition depends on the single-particle wave functions and their overlap being influenced by the number of charge carriers constituting the complex. For equal specific transition times (STTs), the exciton (X) decay time is expected to be twice as long as the biexciton (XX) decay time since the XX has two possible decay channels and the X has only one. 
Previous studies of the decay of X and XX in InAs QDs were limited to a few QDs and reported factors vary from 7.5 \cite{baier06} via 2.3 \cite{thompson01} to 1.1 \cite{zinoni06}. Theoretical predictions range between 4 and 1.5 \cite{wimmer06, narvaez06}.

In this letter, we present a study of 26 different single InAs/GaAs QDs yielding an average factor of $1.7\pm0.4$ for the decay time ratio of X/XX. Furthermore, we expand the study to positively charged excitons (X+) and biexcitons (XX+). We predict a ratio of 1.5 and observe experimental values yielding $1.5\pm0.2$. This close agreement shows that the STTs for X+ and XX+ are very similar in contrast to the STTs of X and XX.

The InAs QDs were grown by metal-organic chemical vapor deposition epitaxy in a GaAs matrix on a GaAs(001) substrate. During QD growth, the rotation of the wafer was interrupted and nominally 1.6 monolayers of InAs were deposited at 500 °C. The interruption leads to a lateral gradient of the QD density. In certain regions of the wafer, the QD density is extremely low ($\approx$10$^7$ cm$^{-2}$) giving spectroscopic access to single QDs.

The QDs were investigated using a JEOL JSM 840 scanning electron microscope equipped with a cathodoluminescence setup \cite{christen91}. The sample was mounted onto the tip of a helium flow cryostat providing temperatures as low as 6~K. All measurements throughout this letter were performed at this temperature. The luminescence was dispersed by a 0.3~m monochromator equipped with a 1200 lines/mm grating. Single QDs were identified by luminescence mapping of the sample. The various QD emission lines were assigned to different excitonic complexes following Ref. \cite{rodt05}.

For time-integrated measurements, a liquid-nitrogen cooled Si charge-coupled-device camera was used. The time-resolved measurements were performed with a Si avalanche photo diode and a beam-blanking unit for pulsed excitation. The beam-on time was 5~ns with a repetition rate of 25~MHz. This pulse configuration allowed luminescence decay from a steady-state situation thus excluding the influence of the capture process. Performing transient line shape analysis, the decay times could be evaluated with an accuracy of 0.1~ns. The dynamic range of these experiments covered up to three orders of magnitude enabling the observation of weak slow components.

Typical transients of the decay of four different excitonic complexes (X, XX, X+, and XX+) in one QD are shown in Fig.~1 \cite{xx+}. All transients are dominated by a fast initial decay typically followed by a second slower component. The decay from different QDs shows a systematic pattern: The XX+ decays the fastest followed by the XX, by the X+, and finally by the X.

The amplitude of the slower component of the transients is very small. For X and X+ its relative amplitude compared to the fast decay ranges between 0 and 0.1 and for XX and XX+, it ranges between 0 and 0.01. The time constants of the slower components vary between 4 and 9~ns. This component originates from processes feeding the initial state of the decay after the external excitation has been switched off. X and X+ having longer initial decay times exhibit a more pronounced second component than XX and XX+, suggesting a feeding process from an external charge carrier reservoir rather than an intrinsic feeding. Such slow components were previously controversially attributed to reemission and lateral transfer of charge carriers \cite{rodt03b, shkolnik05} or to conversion of dark excitons to bright excitons via spin-flip processes \cite{smith05}. For the X+, XX, and XX+ decays, the spin flip can be ruled out as these complexes possess no such dark states. Instead, the source of the feeding process is suggested to be outside of the QDs, e.g., neighboring shallow defect states or reemission from other QDs. In the following, we will focus on the dominant fast component.

\begin{figure}
	\centering
		\includegraphics[width=1\columnwidth]{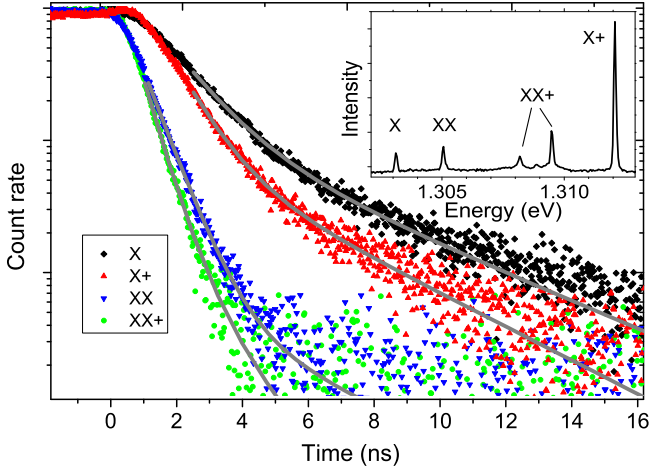}
	\caption{Typical transients of the different excitonic complexes (X, X+, XX, XX+) of one QD. The transients were fitted by a biexponential function. The fast component constitutes the radiative decay time. The second component stems from a refilling process from neighboring defect states or other QDs. The inset shows a typical time-integrated spectrum of a single QD.}
	\label{fig:einQDtr}
\end{figure}

Figure~2 shows the dominant decay times of X, X+, XX, and XX+ as a function of the X+ recombination energy. A clear sequence of decay times for the excitonic complexes is obvious. X exhibits the longest decay time followed by X+, XX, and finally XX+, analogous to Fig.~1 and in agreement with previous reports on InAs/GaAs QDs \cite{thompson01, baier06}. 

No clear trend for a dependence of the decay times on the X+ energy is identified. Apparently, the structural parameters, which lead to differences in the transition energies, do not control the decay times in the same way. Interestingly, the scatter of the X decay times is much larger than that of the other excitonic complexes. The mean decay times are $\overline{\tau(X)}=1.22\pm0.25$~ns, $\overline{\tau(X+)}=0.97\pm0.15$~ns, $\overline{\tau(XX)}=0.76\pm0.12$~ns, and $\overline{\tau(XX+)}=0.66\pm0.14$~ns.  The standard deviation of $\overline{\tau(X)}$ is almost twice as large as for the other complexes. The X wave function seems to be more sensitive to the structural properties of the QD than the wave functions of complexes containing more holes.

The decay times of the neutral and charged complexes will now be compared with respect to the number of allowed transitions of each excitonic complex.

\begin{figure}
	\centering
		\includegraphics[width=1\columnwidth]{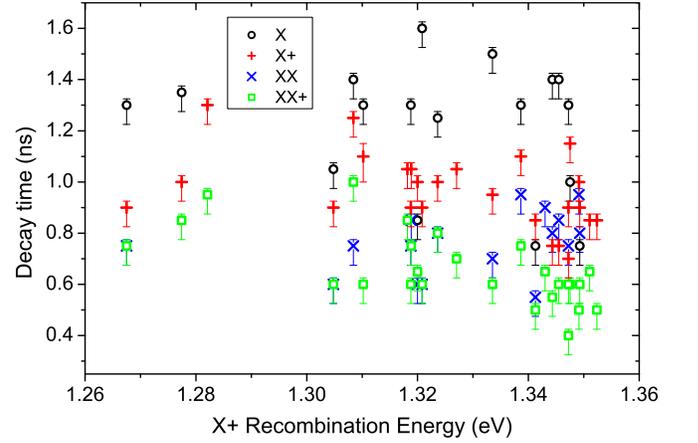}
	\caption{The dominant decay times of the different excitonic complexes are shown as a function of the X+ recombination energy. No clear dependence on energy can be seen.}
	\label{fig:Zeiten-Energie}
\end{figure}

The radiative decay time $\tau(EC)$ of an excitonic complex EC is composed of the specific transition time $\tau_{if}$ of all recombination channels involved \cite{narvaez05}:
\begin{equation}
	\frac{1}{\tau(EC)}=\sum_{i,f}n_{i}\frac{1}{\tau_{i,f}},
	\label{tau}
\end{equation}
with the mean population $n_i$ ($\sum n_i=1$) of the initial state. It is assumed that the STTs for different transition channels ($|i\rangle\rightarrow|f\rangle$ and $|i'\rangle\rightarrow|f'\rangle$) of one complex are identical ($\tau_{i,f}=\tau_{i',f'}=\tau_{EC}$) since the oscillator strength is independent of the actual spin configuration as long as the spin selection rules are obeyed. Under this precondition, the expected excitonic decay times $\tau(X),~\tau(X+),~\tau(XX)$, and $\tau(XX+)$ are determined by the number of allowed recombination channels and their STTs, respectively.

\begin{figure}[b]
	\centering
		\includegraphics[width=1\columnwidth]{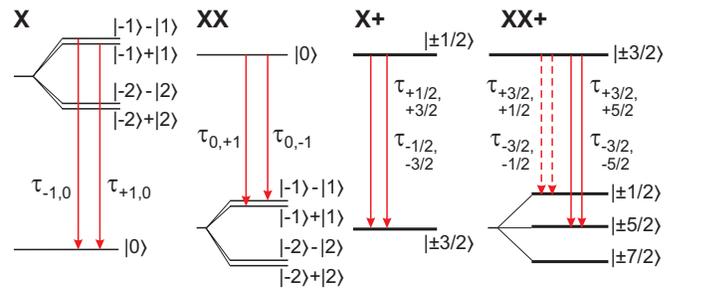}
	\caption{Energy scheme for the recombination channels of the different excitonic complexes. The spin configurations of the different initial and final states are shown. Thick lines correspond to twofold degenerate states.}
	\label{fig:Ubergang}
\end{figure}

Figure 3 shows the energy schemes of the four excitonic complexes to illustrate the different transition channels. The initial and final states are labeled with their respective total spin configurations.

The initial configuration of the optically active X consists of the two bright states with the spin configuration $|-1\rangle\pm|1\rangle$ \cite{bayer02, norm}. The dark states $|-2\rangle\pm|2\rangle$ are optically forbidden and do not contribute to the recombination process. The exciton fine structure splitting of the QDs probed here was determined to be less than 20~$\mu$eV. Hence, the mean initial population $n_{\pm1}$ of the two bright states is $\frac{1}{2}$. Each state possesses one recombination channel to the ground state $|0\rangle$. The decay time then amounts to $\tau(X)=\tau_X$. 

In contrast to X, XX consists of a single initial state $|0\rangle$ with $n_0=1$ and possesses two recombination channels to the two final states: the bright states of the X \mbox{($|-1\rangle\pm|1\rangle$)}. Hence, the decay time amounts to $\tau(XX)=\frac{1}{2}\tau_{XX}$.

To compare the decay times of X and XX, identical STTs for both complexes are assumed. Narvaez \textit{et al.} \cite{narvaez05} calculated almost identical STTs ($\tau_X=\tau_{XX}$) for the X and XX in lens shaped InAs/GaAs QDs of various sizes. Under this assumption, a decay time ratio of $\frac{\tau(X)}{\tau(XX)}=$~2 follows.

\begin{figure}[t]
	\centering
		\includegraphics[width=1\columnwidth]{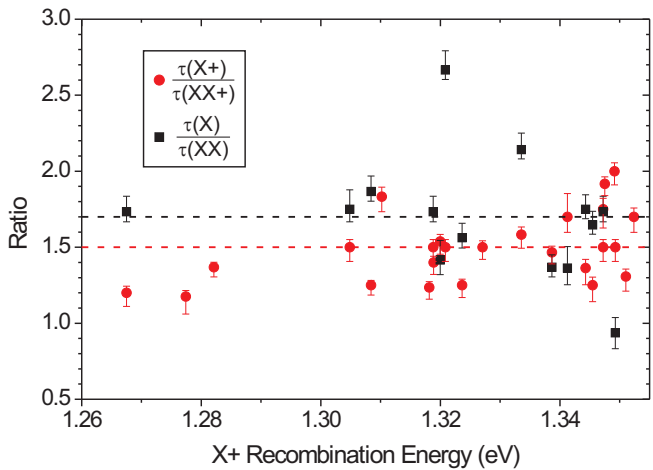}
	\caption{Ratio of the decay times of the neutral and charged complexes as a function of X+ recombination energy. The dashed lines indicate the mean values. The means of the ratios are 1.7$\pm$0.4 and 1.5$\pm$0.2 respectively. The greater scatter of the $\frac{\tau(X)}{\tau(XX)}$ ratio stems from the greater scatter of the X lifetime.}
	\label{fig:Verhaltnis}
\end{figure}

The initial and final states of X+ consist of two degenerate states $|\pm\frac{1}{2}\rangle$ and $|\pm\frac{3}{2}\rangle$, respectively. For each initial state, only one final state is allowed due to the spin selection rules. Since the initial states are equally populated ($n_{\pm1/2}=\frac{1}{2}$), the decay time of X+ following Eq. (\ref{tau}) yields:
\begin{equation}
\frac{1}{\tau(X+)}=\frac{1}{2}\frac{1}{\tau_{+1/2,+3/2}}+\frac{1}{2}\frac{1}{\tau_{-1/2,-3/2}}\approx\frac{1}{\tau_{X+}}
\end{equation}

The initial configuration of XX+ consists of the two degenerate states $|\pm\frac{3}{2}\rangle$. The final state of the XX+ is a triplet, each component consisting of three doublets \cite{seguin06, kavokin03}. The transition to the $|\pm7/2\rangle$ state is forbidden, while the transition to the $|\pm5/2\rangle$ state is allowed. The transition probability into the $|\pm1/2\rangle$ state is reduced to $\frac{1}{2}$ since the final state includes a forbidden transition \cite{akimov05}. Thus, the decay time of the XX+ accounts to 
\begin{eqnarray}
\frac{1}{\tau(XX+)}&=& \frac{1}{2}\left(\frac{1}{2}\frac{1}{\tau_{+3/2,+1/2}}+\frac{1}{2}\frac{1}{\tau_{-3/2,-1/2}}\right)\nonumber
\\ &&+\frac{1}{2}\frac{1}{\tau_{+3/2,+5/2}}+\frac{1}{2}\frac{1}{\tau_{-3/2,-5/2}}\nonumber\\ &\approx&\frac{3}{2}\frac{1}{\tau_{XX+}}
\end{eqnarray}

Just like for the neutral complexes, equal STTs ($\tau_{X+}=\tau_{XX+}$) for the charged complexes are assumed. Since the probed QDs are of small size and, therefore, in the strong confinement regime, adding a neutral pair of charge carriers should alter the wave functions only slightly. The expected ratio of decay times $\frac{\tau(X+)}{\tau(XX+)}$ then amounts to 1.5.

In Fig.~4, the measured ratios of the X and XX decay times (black squares) and of the X+ and XX+ decay times (red circles) are plotted as a function of X+ recombination energy. Again, no apparent trend can be seen. The mean value of $\frac{\tau(X)}{\tau(XX)}$ is 1.7 with a standard deviation of 0.4. For $\frac{\tau(X+)}{\tau(XX+)}$ the mean value results to 1.5 with a standard deviation of 0.2.

The mean $\frac{\tau(X)}{\tau(XX)}$ value is clearly below the predicted value of 2 with a large scatter of the individual values. Qualitatively, the argument based on the number of involved transition channels holds. The exact value, however, crucially depends on the specific QD since the STTs differ a lot from case to case and are not always comparable as in \cite{narvaez05}. Wimmer \textit{et al.}  calculated a size dependent ratio between 1.5 and 2 \cite{wimmer06}. The change in the ratio is governed mainly by $\tau(X)$, thus, supporting our measurements. 

The measured mean value of $\frac{\tau(X+)}{\tau(XX+)}$ matches exactly the expected value of 1.5. The scatter is considerably smaller than for $\frac{\tau(X)}{\tau(XX)}$ with a standard deviation of 0.2 indicating a stronger similarity of the STTs of X+ and XX+ from dot to dot than for the neutral complexes. Here, the number of allowed transitions is the key parameter determining the ratio of decay times of both complexes.

The comparison of the measured $\tau(X)$ and $\tau(X+)$ shows that the STTs $\tau_X$ and $\tau_{X+}$ differ strongly. Adding a single charge carrier to an excitonic complex influences the wave functions and, therefore, their overlap is much stronger than adding a neutral pair of charge carriers. 

In summary, we presented a systematic study of the decay dynamics of excitonic complexes in 26 different single InAs/GaAs QDs. The influence of the specific transition times and electronic fine structure on the dynamics was analyzed. Surprisingly, the scatter of the neutral exciton's decay time from dot to dot is larger than the scatter of the other complexes' decay times indicating a greater sensitivity of the exciton wave function to details of the structural properties of the QD. No dependence on the transition energy was observed in the 1.26 -- 1.35~eV range. The mean value of $\frac{\tau(X)}{\tau(XX)}$ was 1.7 $\pm$ 0.4, thus, below the value of 2 as given by the different number of decay channels alone. This implies shorter STTs for X than for XX. For $\frac{\tau(X+)}{\tau(XX+)}$ a value of 1.5 was predicted. The mean measured ratio is 1.5 $\pm$ 0.2, thus, matching the expected theoretical value. This indicates similar STTs for the transition channels of X+ and XX+. The additional positive charge carrier seems to stabilize the wave function overlap when an additional exciton is added to the complex.

This work was supported by the Deutsche Forschungsgemeinschaft in the framework of SfB 787 and the SANDiE Network of Excellence of the European Commission, Contract No. NMP4-CT-2004-500101. We are indebted to K. Pötschke for the growth of the sample.








\begin{thebibliography}{21}
\bibitem[{xx+()}]{xx+}
\bibinfo{note}{The XX+ transient was always obtained from the more intense of
  the two emission lines. The other line yields the same decay time since the
  occupation of the same complex is probed.}
\bibitem[{nor()}]{norm}
\bibinfo{note}{The normalization factor $\sqrt{2}$ has been disregarded
  throughout this letter for simplicity.}
\end{thebibliography}
\end{document}